\documentclass[aps,prl,twocolumn,showpacs]{revtex4}
\usepackage{graphicx}
\usepackage{dcolumn}

\begin{document}

\preprint{APS/123-QED}

\title{Lattice effects and entropy release at the low-temperature phase transition in the
 spin-liquid candidate $\kappa$-(BEDT-TTF)$_{2}$Cu$_{2}$(CN)$_{3}$}
\author{R. S. Manna$^{1}$}
\author{M. de Souza$^{1}$}
\author{A. Br\"{u}hl$^{1}$}
\author{J.A. Schlueter$^{2}$}
\author{M. Lang$^{1}$}

\address{$^{1}$Physikalisches Institut, J.W. Goethe-Universit\"{a}t
Frankfurt(M), SFB/TR49, D-60438 Frankfurt(M), Germany}
\address{$^{2}$Materials Science Division, Argonne National Laboratory, Argonne, Illinois 60439, USA}


\date{\today}

\begin{abstract}
The spin-liquid candidate
$\kappa$-(BEDT-TTF)$_{2}$Cu$_{2}$(CN)$_{3}$ has been studied by
measuring the uniaxial expansion coefficients $\alpha_{i}$, the specific heat,
and magnetic susceptibility. Special emphasis was placed on
the mysterious anomaly around 6\,K - a potential spin-liquid instability. Distinct and strongly
anisotropic lattice effects have been observed at 6\,K, clearly
identifying this feature as a second-order phase transition. Owing
to the large anomalies in $\alpha_{i}$, the application of
Gr\"{u}neisen scaling has enabled us to determine the
corresponding specific heat contribution and the entropy release. Comparison of the latter with available spin models suggests that spin degrees of freedom
alone cannot account for the phase transition. Scenarios, involving
charge degrees of freedom, are discussed.
\end{abstract}


\maketitle

Organic charge-transfer salts show a rich variety of quantum phases
resulting from the interplay of electronic correlations, low
dimensionality and frustrated magnetic interactions (see
refs.\,\cite{Toyota 07} and \cite{Lebed 08} for recent reviews).
Recently, much excitement has been generated by the proposal of a
quantum spin-liquid (QSL) state in the
$\kappa$-(BEDT-TTF)$_{2}$Cu$_{2}$(CN)$_{3}$ salt \cite{Shimizu 03}.
This system is a half-filled Mott insulator, where
(BEDT-TTF)$_{2}^{+}$ dimers form a quasi-two dimensional (quasi-2D)
triangular lattice characterized by \textit{inter}-dimer hopping
amplitudes $t' \sim t$. The material lacks long-range magnetic order
down to 32\,mK \cite{Shimizu 03}, a small fraction of the estimated
nearest-neighbor Heisenberg exchange coupling $J/k_{B} \simeq$
250\,K \cite{Shimizu 03, Zheng 05}. This observation has stimulated
numerous works aiming at an identification of the QSL state and its
potential instabilities. A number of important questions has been
raised such as the nature of the low-energy excitations
\cite{Yamashita 08, Yamashita 09}, as well as the actual degree of
frustration. According to refs. \cite{Nakamura 09} and \cite{Hem
09}, the system is less frustrated ($t'/t \sim$ 0.8) than was
previously thought \cite{Komatsu 96}, raising the question of how,
under these conditions, a QSL state can be stabilized over magnetic
order \cite{Misguich 99, Morita 02, Zheng 05, Motrunich 05, Yunoki
06, Hayashi 07, Tocchio 09}.

A central issue, which may hold the key for understanding the
puzzling low-$T$ properties, is posed by the mysterious anomaly
around 6\,K, about which very little is known. It appears as a hump
in the specific heat \cite{Yamashita 08}, the NMR relaxation rate
\cite{Shimizu 03, Shimizu 06} and the thermal conductivity
\cite{Yamashita 09}. Various scenarios have been suggested including
a crossover from a thermally to a quantum disordered state
\cite{Yamashita 08} or an instability of the QSL. For the latter, a
spin-chirality ordering \cite{Baskaran 89}, a $Z_{2}$ vortex
formation \cite{Kawamura 84}, a pairing of spinons \cite{Lee 07}, an
exciton condensate \cite{Qi 08}, or a nematic paired state
\cite{Grover 09} have been discussed. Unfortunately, neutron
scattering - the technique of choice for identifying a potential
order parameter - is very difficult to apply due to the organic
nature of the material. However, according to refs. \cite{Lee 07}
and \cite{Qi 08}, the lattice strain could constitute such a
sensitive probe as the QSL instabilities discussed there break the
lattice symmetry.

In this Letter, we report on a study of lattice effects by means of
thermal expansion measurements. Our data reveal a distinct and
strongly anisotropic anomaly at 6\,K - clear evidence for a
second-order phase transition. The application of a thermodynamic
analysis enabled us to determine the entropy release at the
transition. This quantity provides a crucial test for any model attempting
to describe the low-$T$ properties of this material.

The thermal expansion measurements were carried out by using an
ultrahigh-resolution capacitive dilatometer (built after \cite{Pott
83}) enabling the detection of length changes $\Delta l \geq 10^{-2}
{\AA}$. The calorimetric and magnetic measurements were performed
employing a micro-calorimeter \cite{Mueller 02} and a SQUID
magnetometer (Quantum Design MPMS), respectively. The single
crystals used were prepared by following the standard procedure
\cite{Geiser 91}. The crystals, taken from two independently
prepared batches, have the shape of flat distorted hexagons with
dimensions of about 1 $\times$ 0.8 $\times$ 0.1 mm$^3$. The uniaxial
pressure exerted on the crystal by the dilatometer ranges from 1 to
6\,bar.

\begin{figure}[floatfix]
\begin{center}
  \includegraphics[width=0.9\columnwidth]{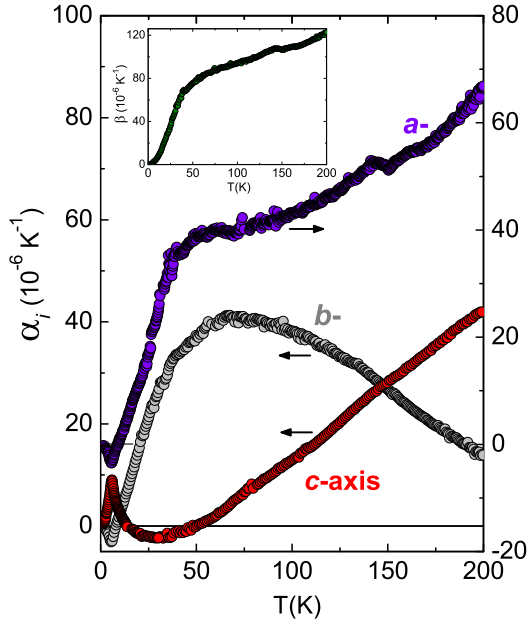}\\[-0.7cm]
   \caption{(Color online) Uniaxial expansivities $\alpha_{i}$ of $\kappa$-(BEDT-TTF)$_{2}$Cu$_{2}$(CN)$_{3}$
   along the in-plane $i = b$ and $c$ axes (left scale) and along the out-of-plane $i = a$ axis (right scale). Inset shows the volume expansion coefficient $\beta$.}
 \label{fig:1}
\end{center}
\end{figure}
\begin{figure}
\includegraphics[width=0.94\columnwidth]{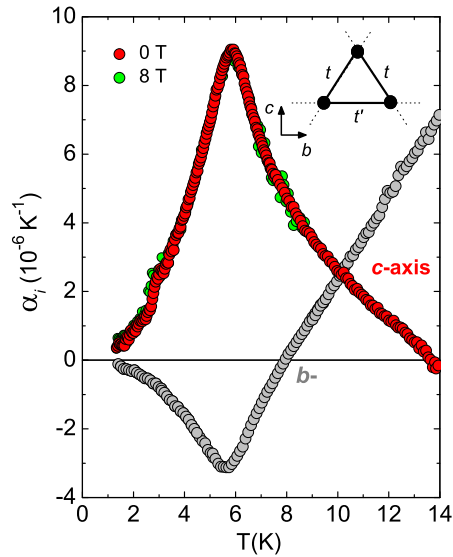}\\[-0.7cm]
\caption{(Color online) In-plane expansivities of
$\kappa$-(BEDT-TTF)$_{2}$Cu$_{2}$(CN)$_{3}$ on expanded scales with
$\alpha_{c}$ taken in $B$ = 0 (red circles) and 8\,T (green circles).
Inset: 2D triangular-lattice dimer model with hopping amplitudes
$t'$ and $t$.} \label{fig:2}
\end{figure}

In Fig.\,\ref{fig:1}, we show the results of the uniaxial expansion
coefficients $\alpha_{i}(T)=l_{i}^{-1}dl_{i}/dT$ measured along the
in-plane $i = b$ and $c$ axes and the cross-plane $a$ axis below
200\,K. Besides the distinct anomalies at low temperatures, which
will be discussed below, the data reveal strongly anisotropic
expansivities. Surprisingly, a particularly strongly pronounced
anisotropy is found for the in-plane data $\alpha_{b}$ and
$\alpha_{c}$, both of which deviate markedly from an ordinary
lattice expansion. Upon cooling, $\alpha_{c}$ becomes reduced almost
linearly down to 70\,K, changes sign at 50\,K and passes through a
shallow minimum around 30\,K. This contrasts with the pronounced
maximum in $\alpha_{b}$ at $T_{max} \sim$ 70\,K. The temperature
range around $T_{max}$ is also distinct in the out-of-plane data
$\alpha_{a}$, where a shoulder appears. The anomalous
$T$-dependencies in $\alpha_{i}$ suggest that, besides phonons,
other excitations contribute significantly to the thermal expansion.
For example, the sign change in $\alpha_{c}$ for $T$ below about
50\,K followed by a minimum indicates a substantial negative
contribution which is maximum around 30\,K. Such a broad negative
anomaly, lacking any signature in the magnetic properties
\cite{Shimizu 03}, may have different origins such as geometrical
frustration and/or quenched disorder (see, e.g.\,\cite{Ramirez 00}).
Likewise, a rounded feature, similar to the one observed in the
magnetic susceptibility $\chi(T)$ at $T_{\chi} \approx 85$\,K
\cite{Shimizu 03}, is expected at $T_{\alpha} \sim T_{\chi}$ as a
result of short-range antiferromagnetic correlations \cite{Bruehl
07}. Irrespective of the nature of the various anomalies in
$\alpha_{i}$, the distinct $\alpha_{b}$ vs.\,$\alpha_{c}$ anisotropy
implies pronounced $T$-dependent in-plane lattice distortions. The
effect is particularly distinct for $T \lesssim$ 50\,K, where upon
cooling the $b$-axis lattice parameter strongly contracts (large
positive $\alpha_{b}$) while the $c$-axis lattice constant expands
($\alpha_{c} <$ 0). Since the hopping amplitudes $t'$ and $t$ depend
sensitively on the lattice parameters, we expect that cooling in
this temperature range is accompanied by an increase of $t'/t$
(cf.\,inset of Fig.\,\ref{fig:2}).

Turning to the anomaly at 6\,K, shown in Fig.\,\ref{fig:2} on
expanded scales, the uniaxial expansivities reveal a distinct peak
of positive ($\alpha_{c}$) and negative ($\alpha_{b}$ and
$\alpha_{a}$, cf. Fig.\,\ref{fig:1}) sign, which is most strongly
pronounced in $\alpha_{c}$. The shape of this feature and its
sharpness are clear indications of a phase transition, albeit of
distinctly non-mean-field type \cite{comment1}, signalizing the
presence of strong critical fluctuations (see, e.g. \cite{de Souza
07, de Souza 08}). Measurements taken upon cooling and warming at a
very slow rate of $\pm$ 1.5 K/h failed to detect any hysteresis,
consistent with a second-order transition. Around 3\,K the
$\alpha_{c}$ data reveal indications for yet another anomaly of much
smaller size, reproducible in detail in consecutive runs. We stress
that the features at 6\,K and 3\,K in $\alpha_{c}$ remain unaffected
by a magnetic field of 8\,T applied along the $c$-axis
(cf.\,Fig.\,\ref{fig:2}). Remarkably enough, the volume expansion
coefficient $\beta = \alpha_{a} + \alpha_{b} + \alpha_{c}$ shows a
much less peculiar behavior (cf. inset of Fig.\,\ref{fig:1}). In
particular, $\beta$ varies smoothly around 50\,K and lacks any
anomaly at 6\,K, i.e., $\Delta \beta\mid_{6K} \approx$ 0. According
to the Ehrenfest relation, this implies that the 6\,K transition is
practically insensitive to hydrostatic pressure.

\begin{figure}
\includegraphics[width=1.0\columnwidth]{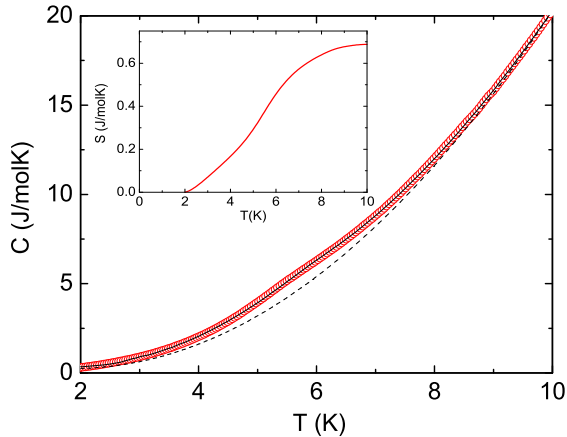}\\[-0.5cm]
\caption{(Color online) Specific heat of
$\kappa$-(BEDT-TTF)$_{2}$Cu$_{2}$(CN)$_{3}$ between 2 and 10\,K.
Solid line represents the result of a least-squares fit of $C_{fit}(T)$
(see text), yielding a smooth background $C^{bg}$ (broken line). Inset:
entropy $S^{an}$ obtained by integrating $\delta C^{an}/T$ from 2\,K to 10\,K.} \label{fig:3}
\end{figure}

\begin{figure}
\includegraphics[width=0.8\columnwidth]{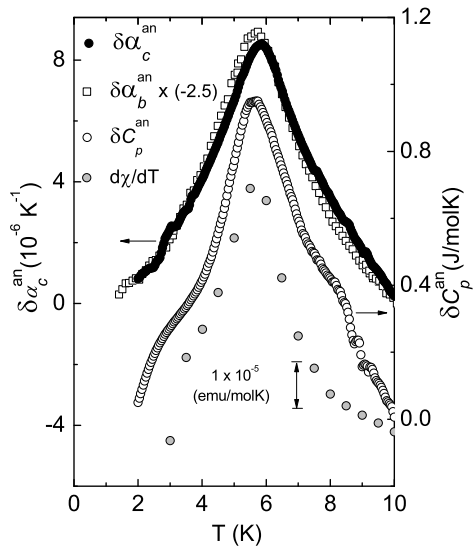}\\[-0.5cm]
\caption{(Color online) Phase transition anomalies, derived as described in the text, in $\alpha_{c}$
and $\alpha_{b}$ (multiplied by a factor $-$2.5) (left scale), $C$ (right scale), and $d\chi/dT$ (scale is given by the arrow).} \label{fig:4}
\end{figure}

In Fig.\,\ref{fig:3} we show results of the specific heat on a small
single crystal of mass $\lesssim$ 100\,$\mu$g for temperatures 2\,K
$\leq T \leq$ 10\,K. The data reveal a smooth increase with $T$ and
a hump-like feature around 6\,K, consistent with literature results
\cite{Yamashita 08}. Evidently, the quantity of interest - the
contribution associated with the phase transition $\delta C^{trans}$
- is difficult to separate from the background $C^{bg}$. Although
$C^{bg}$ is likely to be dominated by the lattice specific heat
$C^{ph}$ (unfortunately unknown) it may also contain other
contributions. Attempts to separate $\delta C^{trans}$ by
subtracting from the measured specific heat $C^{ph}$ of a related
salt with a different anion \cite{Yamashita 08}, involve
considerable uncertainties. In order to overcome this problem, we
use an Ansatz, which has proved particularly valuable for analyzing
phase transitions in organic materials \cite{de Souza 09}. The
approach is based on the assumption of a proportionality between
corresponding contributions to the expansivity and specific heat
$\delta \alpha_{i} \propto \delta C$. This so-called
Gr\"{u}neisen-scaling applies provided that there is only one
relevant energy scale in the $T$ range of interest (see \cite{Bruehl
07} and references cited therein). In what follows, we focus on the
range 2\,K $\leq T \leq$ 10\,K, where this assumption should be
valid. First, the data in Fig.\,\ref{fig:2} were used to extract the
anomalous contribution $\delta \alpha_{c}^{an}$ associated with the
peak anomaly around 6\,K. To this end, smooth background curves were
subtracted from $\alpha_{c}$ and $\alpha_{b}$ such that the
remaining contributions show the scaling behavior $\delta
\alpha_{c}^{an} \propto \delta \alpha_{b}^{an}$ required by
thermodynamics (cf. Fig.\,\ref{fig:4}). Note that due to the large
size of the anomaly in $\alpha_{c}$, uncertainties in the background
have only little effect on the resulting $\delta\alpha_{c}^{an}$.
With the so-derived $\delta\alpha_{c}^{an}$ a least-squares fit to
the specific heat was performed using the function $C_{fit}(T)=
\delta C^{an} + C^{bg} = \frac{1}{\gamma_{c}}\delta\alpha_{c}^{an}+
C^{bg}$, with a fit parameter $\gamma_{c}$, representing a
generalized Gr\"{u}neisen parameter \cite{Bruehl 07}. A
parameterization of the form $C^{bg} = a_{0} + a_{1}T + a_{2}T^{2}$
was used. The good quality of the fit, yielding $\delta C^{an}$ (cf.
Fig.\,\ref{fig:4}) and a smooth background function $C^{bg}$ (broken
line in Fig.\,\ref{fig:3}), is reflected by the small residual
$(C(T) - C_{fit}(T))/C(T)$ of less than 1.5 \% in the range 4\,K
$\leq T \leq$ 10\,K.

The notion of a second-order phase transition at 6\,K is also
corroborated by reassessing the low-$T$ magnetic susceptibility
$\chi(T)$. The results of our experiment, being in good overall
agreement with published data \cite{Shimizu 03}, are plotted in
Fig.\,\ref{fig:4} as $d\chi/dT$ vs.\,$T$. The measurements were
performed by using a composite sample consisting of a large number
of randomly oriented single crystals of total mass of 18.0 mg. As
the figure indicates, $d\chi/dT$ shows a distinct peak, the shape
and position of which practically coincide with the phase transition
anomalies in $\alpha_{i}(T)$ and $C(T)$. Such a conformity between
$d\chi/dT$ and thermodynamic quantities, known from magnetic
transitions \cite{Fisher 62}, indicates that spin degrees of freedom
are involved to some extent in the transition. However, as will be
shown below, their contribution alone appears insufficient to
account for the observed entropy release.

By integrating $\delta C^{an}/T$ from 2\,K up to 10\,K, we obtain an
entropy $S^{an}$ of (0.69 $\pm$ 0.05)\,J\,mol$^{-1}$\,K$^{-1}$ (cf.
inset of Fig.\,\ref{fig:3}), with one mole substance containing
2$\cdot N_{A}$ dimers (spins) and $N_{A}$ denoting Avogadro's
constant. This represents a lower bound of the true entropy
$S^{trans}$ associated with the phase transition, see below. The
so-derived $S^{an}$ corresponds to about (6 $\pm$ 0.4)\% of the
system's full spin entropy (this would be $R$ln2 for a hypothetical
system containing $N_{A}$ spins with $S = 1/2$, with $R$ the
universal gas constant). It can be compared with calculations of the
magnetic entropy for 2D triangular-lattice $S$ = 1/2 systems,
unfortunately available only for the isotropic ($t'$ = $t$) case.
For the pure Heisenberg antiferromagnet \cite{Bernu 01}, which may
serve as a rough estimate, we find a residual entropy for
temperatures $k_{B}T \leq$ 0.04$\cdot J$, corresponding to $T \leq$
10\,K for $J/k_{B}$ = 250\,K, of only (2.3 $\pm$ 0.1) \% of $R$ln2.
By including ring exchange processes, likely to be significant near
the Mott transition and which may account for the absence of
N\'{e}el ordering \cite{Misguich 99}, a somewhat larger value is
expected. According to a variational study \cite{Motrunich 05},
these ring-exchange processes give rise to a low-$T$ QSL state
characterized by gapless fermionic excitations. This implies a
spinon contribution to the specific heat of $C_{spinon}$ =
$\gamma_{spinon}\cdot T$, with $\gamma_{spinon}$ =
($\pi^{2}$/3)$k_{B}^{2}N_{A}n(E_{F})$ and a spinon density of states
at the "Fermi surface" $n(E_{F})$ = 0.28/$t_{spinon}$
\cite{Motrunich 05}. For $t_{spinon}$ = $J$ \cite{Motrunich 05}, we
find $\gamma_{spinon}$ = 30 mJ mol$^{-1}$K$^{-2}$ and a remaining
spinon entropy of 5.2 \% of $R$ln2 for $T \leq$ 10\,K. In
Ref.\,\cite{Grover 09} it has been argued that this QSL state is
instable against the formation of spinon pairs - a nematic QSL state
- by which $S^{spinon}$ is released to a full extent or partly (in
the presence of impurity scattering).

This residual spinon entropy $S^{spinon}$ appears to be quite close
to the experimentally determined entropy release $S^{an}$, seemingly
consistent with the scenario discussed in ref.\,\cite{Grover 09}.
However, by estimating possible errors, the discrepancy between
$S^{spinon}$ and $S^{an}$ becomes significant. First, we note that
the above-mentioned underestimation of $S^{trans}$ by $S^{an}$ is
considerably large for the present case. The reason for this is
mainly due to the peculiar non-mean-field shape of the transition,
which required a truncation of the high- and low-$T$ flanks in
$\delta C^{an}$ (cf. Fig.\,\ref{fig:4}). Simulations for quantifying
the associated error indicate that $S^{trans}/S^{an}$ is probably in
the range 1.2 - 1.4. Moreover, we expect that the available low-$T$
spin entropy will be further reduced for an anisotropic triangular
lattice ($t' < t$) - very likely the case for the present material
\cite{Nakamura 09, Hem 09}. Thus, an estimated entropy release
$S^{trans}$ of about 7-8 \% of $R$ln2 has to be contrasted with an
available spin entropy of less than 5.2 \% of $R$ln2. We consider
this mismatch of about 30 - 50\%, by which the experimental value
exceeds $S^{spinon}$, to be a strong indication that spin degrees of
freedom alone are insufficient to account for the phase transition.
An obvious candidate to rationalize the mismatch would be charge
degrees of freedom. In fact, according to optical conductivity
measurements \cite{Kezsmarki 06}, the charge-gap in this material is
strongly suppressed indicating the material's close proximity to the
insulator-metal transition.

As a possible scenario we mention the proposed formation of an
exciton condensate \cite{Qi 08}. This is accompanied by a charge
modulation breaking lattice symmetry. Objections to this scenario
may be due to the lack of clear signatures for a charge modulation
below 6\,K in $^{13}$C-NMR studies \cite{Shimizu 06}. The poverty of
evidence, however, might be due to the anomalous inhomogeneous
broadening of the NMR spectra \cite{Shimizu 06}, possibly obscuring
a potential line splitting.

Alternatively, one might consider charge fluctuations at elevated
temperatures, causing fluctuating electrical dipoles. Some evidence
in support of this has been recently observed in dielectric
measurements \cite{Abdel 09}. Within such a scenario, the 6\,K
transition might indicate an ordering of those preformed dipoles. We
stress that a distinct type of charge ordering, driven by an
increase in the degree of frustration $t'/t$, has recently been
predicted for the present material \cite{Li 09}. Due to the ionic
nature of the material, ordering phenomena in the charge sector are
expected to give rise to distinct lattice effects (see, e.g.
\cite{de Souza 08, de Souza 09}). Since the charge distribution also
affects the magnetic exchange constants via $J \propto t^{2}/U$ with
$U$ the on-site Coulomb repulsion, a response in the magnetic
susceptibility is also expected.

In summary, measurements of the uniaxial expansivities on the
spin-liquid candidate $\kappa$-(BEDT-TTF)$_{2}$Cu$_{2}$(CN)$_{3}$ yield distinct and strongly
anisotropic lattice effects at 6\,K, clearly
identifying this phenomenon as a second-order phase transition. By
means of a Gr\"{u}neisen-scaling Ansatz, the corresponding anomaly
in the specific heat could be separated. Comparison of the associated entropy
release with spin models suggests that charge
degrees of freedom are involved in the transition.

\begin{acknowledgments}
We acknowledge discussions with T. Sasaki, R. Valent\'i and C. Gros. Work supported by Argonne, a U.S. Department of Energy Office of Science laboratory, operated under Contract No. DE-AC02-06CH11357.
\end{acknowledgments}


\begin{thebibliography}{Buravov 271}
\bibitem{comment1} We exclude the possibility of a magnetic transition induced by sample inhomogeneities: no signatures for magnetic order have been seen by NMR studies \cite{Shimizu 03, Shimizu 06}; no sample-to-sample variation has been found for the transition temperature for the various crystals studied here.
\end{thebibliography}
\end{document}